Magnetic imaging and domain nucleation in CrSBr down to the 2D limit.

Yishay Zur[†,1,2], Avia Noah[†*,1,2], Carla Boix-Constant[3], Samuel Mañas-Valero*[3], Nofar Fridman[1,2], Ricardo Rama-Eiroa[4,5], Martin E. Huber[6], Elton J. G. Santos*[5,7,4], Eugenio Coronado[3], and Yonathan Anahory*[1,2]

[†] These authors contributed equally to this work.
[1] The Racah Institute of Physics, The Hebrew University, Jerusalem, 9190401, Israel
[2] Center for Nanoscience and Nanotechnology, Hebrew University of Jerusalem, Jerusalem, 91904, Israel
[3] Instituto de Ciencia Molecular (ICMol), Universitat de València, Catedrático José Beltrán 2, Paterna 46980, Spain
[4] Donostia International Physics Center (DIPC), 20018 Donostia-San Sebastián, Basque Country, Spain
[5] Institute for Condensed Matter Physics and Complex Systems, School of Physics and Astronomy, University of Edinburgh, Edinburgh, EH93FD, United Kingdom
[6] Departments of Physics and Electrical Engineering, University of Colorado Denver, Denver, CO 80217, USA
[7] Higgs Centre for Theoretical Physics, University of Edinburgh, Edinburgh EH93FD, United Kingdom

*Correspondences to: avia.noah@mail.huji.ac.il, Samuel.Manas@uv.es, esantos@ed.ac.uk, yonathan.anahory@mail.huji.ac.il

## Abstract

Recent advancements in 2D materials have revealed the potential of van der Waals magnets, and specifically of their magnetic anisotropy that allows applications down to the 2D limit. Among these materials, CrSBr has emerged as a promising candidate, because its intriguing magnetic and electronic properties have appeal for both fundamental and applied research in spintronics or magnonics. Here, nano SQUID-on-tip (SOT) microscopy is used to obtain direct magnetic imaging of CrSBr flakes with thicknesses ranging from monolayer ($N$=1) to few-layer ($N$=5). The ferromagnetic order is preserved down to the monolayer, while the antiferromagnetic coupling of the layers starts from the bilayer case. For odd layers, at zero applied magnetic field, the stray field resulting from the uncompensated layer is directly imaged. The progressive spin reorientation along the out-of-plane direction (hard axis) is also measured with a finite applied magnetic field, allowing to evaluate the anisotropy constant, which remains stable down to the monolayer and is close to the bulk value. Finally, by selecting the applied magnetic field protocol, the formation of Néel magnetic domain walls is observed down to the single layer limit.

## Introduction

Magnetic anisotropy is an essential parameter that governs the magnetic properties of low dimensional materials and has attracted much interest since the discovery of magnetically ordered van der Waals (vdW) materials.[1–3] In fact, crystalline anisotropies contribute to long-range magnetic order,[4–7] determine the magnetic domain wall structure,[8–11] and give rise to exotic spin textures.[12,13] Thus, understanding the underlying physical mechanisms that control magnetism in these low dimensional materials might allow us to harness their potential applications. For example, such progress could increase the ordering temperature of ultra-confined magnets. In addition to such applications, magnetic anisotropy engineering opens opportunities for interesting experimental realizations in two-dimensions, such as the transverse Ising model,[14] or the formation of non-colinear spin textures such as vortices or skyrmions.[12]

As an example of a layered vdW magnet, CrSBr has recently attracted particular attention due to its tantalizing optical, electronic, and magnetic properties, its potential impact on optoelectronics, spintronics, or magnonics.[15,16] Bulk CrSBr is a type-A antiferromagnet (AFM) formed by ferromagnetic (FM) layers ($ab$ plane) with spins confined in plane (IP) and pointing along the $b$ axis – easy magnetic axis (**Fig. 1a**).[17] Below the Néel temperature (132 K), these layers are coupled antiferromagnetically along the out-of-plane (OOP) direction ($c$ axis). First-principle calculations[18,19] and experimental results[20] indicate that the interlayer AFM coupling is three orders of magnitude weaker than the intralayer FM coupling. As a first approximation, CrSBr is therefore typically modeled as a stack of FM layers with triaxial anisotropy and with easy, intermediate, and hard magnetic axes along the $b$, $a$, and $c$ directions, respectively.[21] At low temperatures, the spins can be reoriented from the $b$ axis to the $a$ and $c$ axis by applying 1 T, and less than 2 T magnetic fields along these axes, respectively.[22,23] The results do not exhibit strong shape dependence, suggesting that the shape anisotropy due to the dipolar field is negligible in this material.

Despite the weak interlayer coupling, the magnetic properties of the monolayer sharply differ from those of the bilayer and multilayer samples, as evidenced indirectly by the magneto-resistance[21,22] (MR) and optical properties.[7] For example, monolayer CrSBr exhibits a positive MR in contrast to the negative MR observed in bilayers and multilayers.[22–24] Little is known about the magnetic properties of monolayer and few-layer CrSBr, mainly due to the scarcity of appropriate experimental tools. Unlike CrI$_3$, where the magnetic moments are oriented OOP, the IP magnetization of CrSBr poses challenges for probing using optical techniques.[7] Moreover, imaging technique such as magnetic force microscopy cannot resolve the static magnetization due to the tip-sample interactions.[25] Therefore, sensitive imaging techniques with negligible back action are necessary to resolve key open questions, such as the preservation of 2D long-range order, the resolution of magnetic textures that these layers may exhibit, or to examine their response to an applied field. Here, we overcome these problems and report magnetic imaging of CrSBr down to the monolayer and bilayer case by scanning SQUID-on-tip (SOT) microscopy. This allows us to image the local stray field emanating from CrSBr flakes with thicknesses ranging from monolayer, $N=1$, to few-layer, $N=5$, when a magnetic field along the hard axis, $c$, is applied, and to resolve the magnetic moment with high spatial resolution. The results reveal intralayer FM order in all thicknesses. For odd layers, at zero applied magnetic field, we directly image the stray field resulting from the uncompensated layer. In addition, at a finite applied magnetic field, we measure a progressive spin reorientation from $b$ to $c$, which allows us to evaluate the corresponding OOP anisotropy constant for different thickness. Surprisingly, this anisotropy is not affected by the thickness and remains unchanged down to the monolayer. Finally, at higher fields, we observe the formation of magnetic domain walls for odd numbers of layers. These results therefore describe the interplay between magnetic anisotropy, magnetic domains, and dimensionality.

# Results

CrSBr flakes with various thicknesses were mechanically exfoliated and, unless otherwise described, were encapsulated with h-BN thin flakes (see Experimental Section). Optical images of the flakes measured in this work are shown in **Figure 1b-e**. The flakes, cleaved along the $ab$ plane, exhibit a rectangular shape, where the short edge ($b$-axis) is typically ten times shorter than the long side ($a$-axis). We note that the easy magnetic axis is parallel to $b$, despite the shape anisotropy that would favor the $a$-axis. The samples were maintained in an inert atmosphere during fabrication and storage but were exposed to air for a few hours while being loaded into the SOT microscope. Each sample was cooled repeatedly.

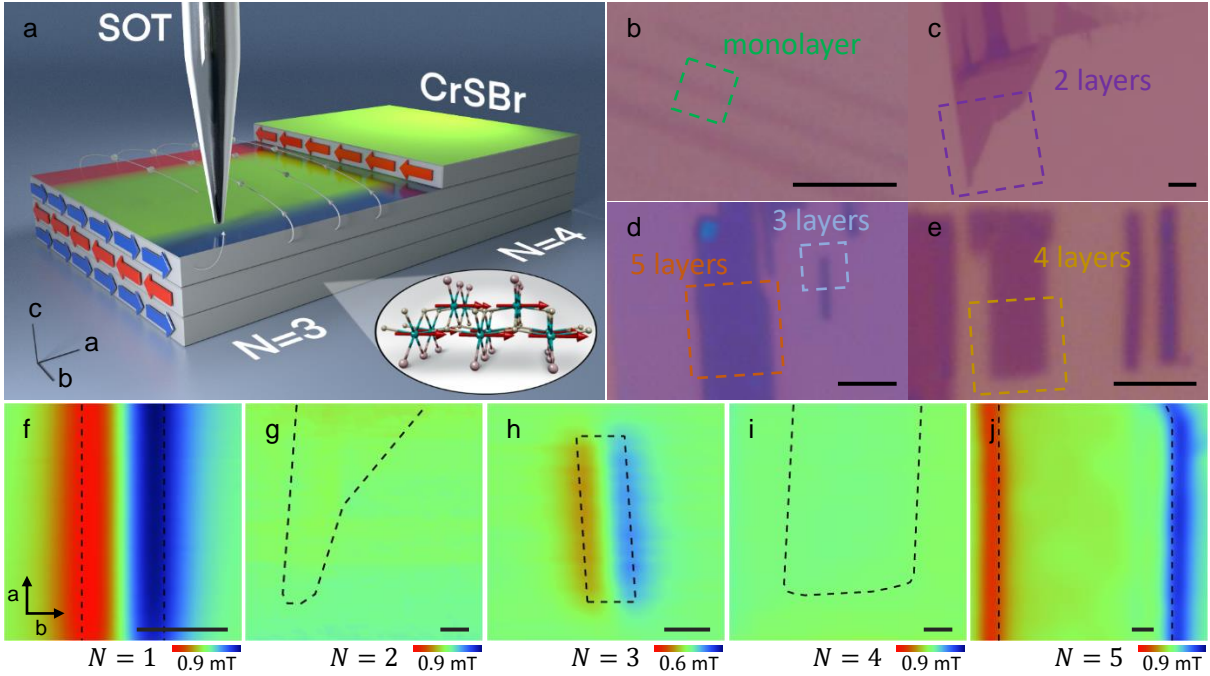

**Figure 1 SQUID-on-tip (SOT) images of CrSBr at 4.2K.** (**a**) Schematic illustration of the measurement setup with a sample with an odd and even number of layers. (**b-e**) Optical images of CrSBr flakes studied in this work ranging from $N=1$ to $N=5$ layers. Panels **b**, **c** and **e** present CrSBr flakes prior to hBN encapsulation, for clarity, while the image shown in **d** was taken after capping with a thin hBN layer. The SOT point of view is marked with dashed lines. The scale bar is 2.5 μm for all optical images. (**f-j**) $B_z(x,y)$ SOT images of CrSBr flakes for $N=1$ to $N=5$ layers. The images are acquired at zero applied field after exposing the sample to an out-of-plane field $|\mu_0 H_z^{exc}| > 1.5$ T. The flake edges are marked with black dashed lines. The scale bar is 400 nm for all SOT images. The red to blue color scale represents lower and higher magnetic fields, with a scale of $B_z=0.9$ mT (**f,g,i,j**) and of $B_z=0.6$ mT (**h**). The tip-to-sample distance was 50 nm **f**, and 200 nm **g, h, I, j**.

The magnetic properties of CrSBr flakes were examined by imaging the local OOP component of the magnetic field $B_z(x,y)$ with a scanning SOT microscope at 4.2 K (**Fig. 1a**, Experimental Section, and Refs. [26,27]). **Figure 1f-j** shows $B_z(x,y)$ images for distinct sample thicknesses at zero applied magnetic field ($\mu_0 H_z=0$) after exposing the sample to a strong OOP magnetic field $|\mu_0 H_z^{exc}| > 1.5$ T. No measurable signal is observed in any of the samples with an even number of layers (**Fig. 1g,i**), which is consistent with a fully compensated type-A AFM as schematized in **Figure 1a** for $N=4$. In contrast, all the samples with an odd number of layers (**Fig. 1f,h,j**), exhibit a magnetic stray field along the long edges of the sample with no measurable field detected along the short edges or in the interior of the sample (as schematized in **Figure 1a** for $N=3$). Opposing long edges have OOP fields $B_z$ with equal magnitude but opposite sign. This feature is visible in the SOT images where the field points OOP along the $c^+$ direction (color-coded in blue) on one edge, and along the $c^-$ direction (color-coded in red) on the opposite side. The SOT images demonstrate a net uniform magnetization parallel to the $b$-axis, which is consistent with a type-A AFM with an uncompensated monolayer. The difference in the magnitude of the measured magnetic signal is explained by the difference in tip-to-sample distance and the lateral dimensions of the flake. Notably, our observations image the magnetic stray fields arising from a single layer directly, thus providing the ability to determine the magnetic order.

As the next step, we examined the thickness-dependence of the magnetization after application of an external magnetic field along the $c$ axis, $H_z$. **Figure 2a-o** presents a set of SOT images at $\mu_0H_z$=-0.5, -0.3, 0, 0.3, and 0.5 T for even and odd number of layers, and monolayer CrSBr, after an initial saturation of the magnetization at larger fields ($|\mu_0H_z^{exc}|$>1.5 T). A uniform stray field $B_z(x,y)$ pointing in the direction of the applied field could be detected above the entire flake in samples with even layers and $H_z\neq0$ (**Fig. 2a-e**). The absence of an antisymmetric signal at the sample edges demonstrates the lack of a net IP magnetization as a result of the interlayer antiferromagnetic coupling. The growth in signal with the field magnitude reflects an increase in OOP moment due to spin canting that reduces the magnetic moment compensation between the layers (**Fig. 2q**). Thus, the net moment detected is not conserved as a function of the applied field. For odd layers (**Fig. 2f-j**), the observed magnetic field distribution is the superposition of the $B_z(x,y)$ image resulting from canted spins (as seen in even layer flakes) and the $B_z(x,y)$ image, which represents the IP projection of the uncompensated layer (as seen in **Fig. 1h**). With growing $H_z$, the IP component of the magnetization decreases at the expense of the OOP component (see **Fig. 2f,j,k,o**). This results in an imbalance between the measured field at each edge. This difference is more evident in the magnetic profile shown in **Supplementary Figure 1**. At fields greater than the saturation field $H_s\lesssim2$ T, the images for odd and even layer samples should be identical when normalized by the number of layers. Our results indicate a similar response of the magnetization down to the monolayer (**Fig. 2k-o** and **Supplementary Figure 2**).

To quantify the change in net magnetization $\mu_{tot}$ in response to an OOP field ($H_z$) we carried out magnetostatic simulations. Based on our SOT images included in **Fig. 2a-o**, we consider a uniform net magnetization restricted in the $bc$ plane. The simulation calculates the magnetic field resulting from a flake with a particular geometry and spin orientation and returns the parameters that minimize the residuals relative to the SOT image. The fitting parameters are the net magnetic moment per unit area $\mu_{tot}/A$, the spin polar angle θ (between $b$-axis to $c$ axis, see **Fig. 2q**), and the location of the flake edges. The tip-to-sample distance (50 – 300 nm) was obtained by coupling the tip to a tuning fork to sense the surface.[28] The flake edges are marked with dashed lines (**Fig 2c,h,m**). The results indicate that the simulated field distribution is in good agreement with the measured SOT image (**Supplementary Figures 3-4**).

**Fig. 2p** presents the $\mu_{tot}$ obtained from the best fit between our simulation and experimental data. As discussed above, $\mu_{tot}/A(H_z=0)\approx0$ for even layers, since all the moments are compensated (**Fig. 2p**, bottom panel). For odd layers with $N>1$ (top panel), $\mu_{tot}/A(H_z=0)$=37±4 $\mu_B$ nm$^{-2}$, which is in excellent agreement with the expected moment for the uncompensated layer ($\mu_{tot,N=1}/A$=36 $\mu_B$ nm$^{-2}$).[17,20,29] For the few-layer case ($N>1$), $\mu_{tot}$ increases parabolically with $H_z$ (**Fig. 2p**), independently of the number of layers. However, for $N$=1, in the absence of moment compensation, $\mu_{tot}$ is conserved. As expected, our results indicate that for $N$=1, the magnetic moment $\mu_{tot}$=33±3 $\mu_B$ nm$^{-2}$ is constant for all measured fields. This value is slightly smaller than the bulk and the few-layer flakes, but within our uncertainty of 10%.

To further characterize the magnetic response, we evaluate the canting angle $\alpha$ of the spins within each layer based on the best fit for the polar angle θ and $\mu_{tot}$ obtained from our simulations. We assume that the IP component of neighboring layers points in opposite directions while preserving the same canting angle $\alpha$, as depicted in **Fig. 2q**, and deduce that $\sin\alpha=\mu_{tot}\sin\theta/N\mu_{(N=1)}$. The results as presented in **Figure 2r**, indicate that the angle $\alpha$ exhibits a similar linear $H_z$ dependence for all measured thicknesses in the explored range of magnetic fields, suggesting that the magnetization is unaffected by the thickness of the sample down to the monolayer.

We complete this analysis by estimating the magnetic anisotropy constant $K_e$ for different layers. For this purpose, we model the relation between $K_e$ and $\alpha$, while assuming that the magnitude of the spin torque ($\tau$) resulting from the Zeeman coupling to the applied field is equal in magnitude to the restoring force resulting from the anisotropy. The standard expression for the spin torque is $|\tau|=|\mu_0H_z\times\mu_{Cr}|=|\mu_0H_z||\mu_{Cr}||\sin(90-\alpha)|$, while we can write the anisotropy restoring torque as $|\tau_a|=K_e|\sin2\alpha|$.[30] Thus, by equating these terms, we obtain $K_e=|\mu_0H_z||\mu_{Cr}||\sin(90-\alpha)|/|\sin2\alpha|$. Surprisingly, the anisotropy is independent of the samples thickness, with $K_e$=0.13±0.03 meV per Cr atom for all layers, a value that is larger than the numerical estimates.[18,19] This value can be compared to the saturation field $H_s$ by neglecting the effect of the magnetic domains, to give $H_s=2K_e/M_s$=1.6 T, which is in good agreement with the reported values in the bulk.[17] In conclusion, our results demonstrate that CrSBr retains its bulk magnetic properties (anisotropy and magnetization) down to the monolayer. Applying the same analysis to a monolayer that was not encapsulated and was exposed to air for some hours (see **Supplementary Figure 5**), yielded a negligible increase of the magnetic moment (36±3 $\mu_B$ nm$^{-2}$) with respect of our uncertainty and a 30% reduction in the magnetic anisotropy (0.1 meV per Cr). These variations are at the limit of experimental uncertainty and indicate that CrSBr is far more stable than other 2D magnets, such as CrI$_3$, although we cannot completely exclude the possibility of some small degradation in the monolayer due to exposure to air.[21]

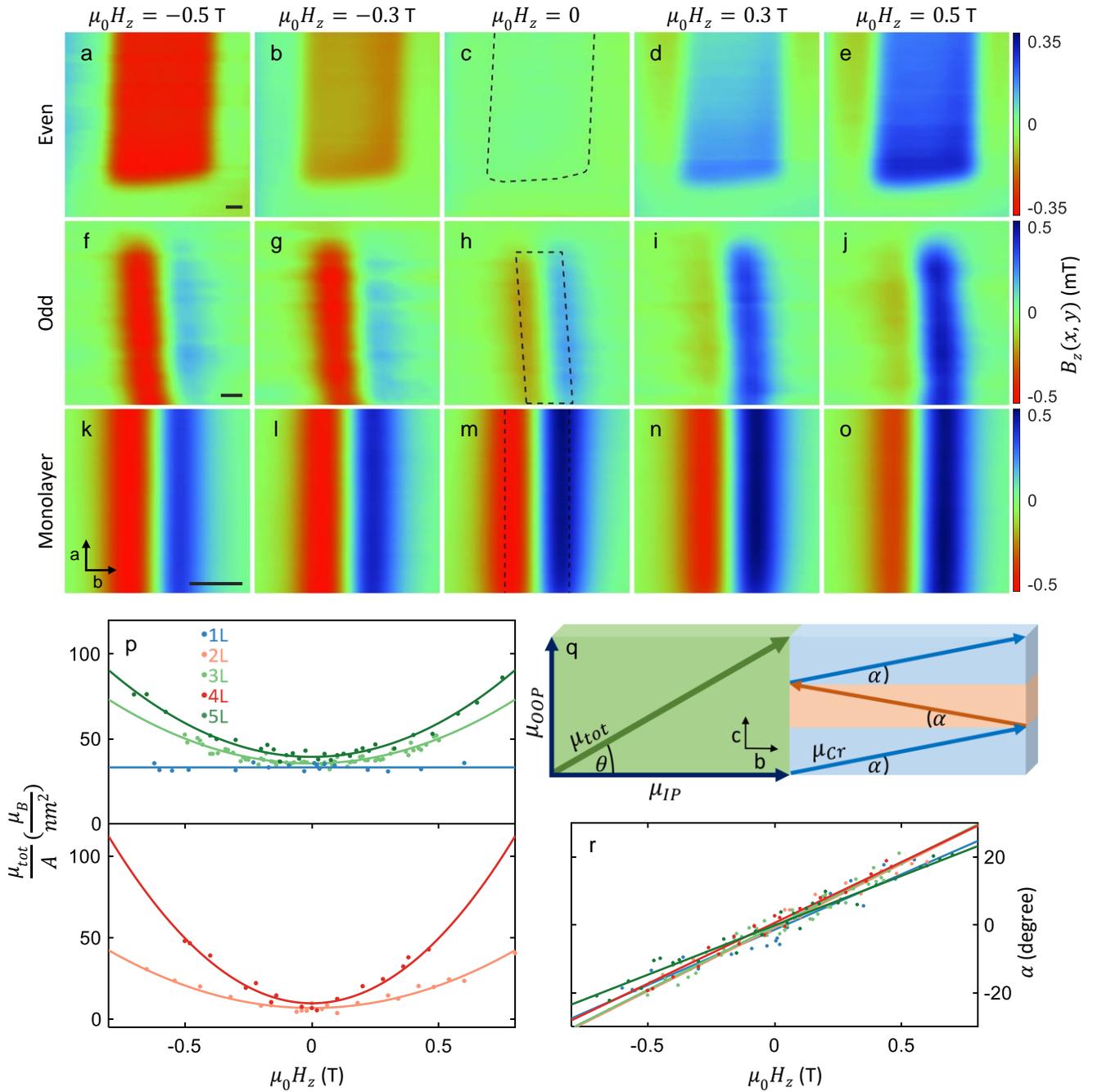

**Figure 2 High anisotropy in a CrSBr monolayer compared to a few-layer sample. (a-o)** Sequence of magnetic SOT images of the OOP component of the local magnetic field $B_z(x,y)$ at $\mu_0H_z$=-0.5, -0.3, 0, 0.3, and 0.5 T after an initial saturation of the magnetization at negative fields $|\mu_0H_z^{exc}|>1.5$ T. Even layers, $N=4$ **a-e**, odd layers, $N=3$ **f-j**, and a monolayer **k-o**. **(p,r)** Magnetostatic simulation results of the net magnetic moment density $\mu_{tot}/A$ **(p)**, and the canting angle $\alpha$ between the $b$ and $c$ axis **(r)**. The anisotropy constant $K_e$ is remarkably constant (up to 30%) down to the monolayer. **(q)** Geometrical relations between the simulation parameters $\mu_{tot}$ and $\theta$ and microscopic parameter $\alpha$ for $N=3$. The scale bar is 300 nm for all images **(a-o)**. The red to blue color scale represents lower and higher magnetic fields, respectively, with a shared scale for each flake.

In order to investigate the magnetic moment response to a larger applied magnetic field in the range 1 T<$|\mu_0 H_z|$<3 T (field required for the spin reorientation along the *c* axis[22,23]), we applied a field excursion at high fields $\mu_0 H_z^{exc} \approx 2$ T and acquired the SOT image $B_z(x,y)$ near zero field at $|\mu_0 H_z|$<30 mT. For odd layers, the saturation of the magnetization at such fields has the effect of changing the direction of the net IP magnetization (compare **Fig. 3a** and **3c**). The apparent coupling between the OOP field and the IP magnetization is most likely due to a small misalignment between the normal of the sample and the axis of the solenoid generating the field. We estimate this tilt to be less than 1°. This tilt implies that the IP projection of $H_z$ = 2 T can reach 35 mT. That value is an order of magnitude smaller than the saturation field along the *b*-axis ($H_s^b$=500 mT). However, at $\mu_0 H_z$ = 2 T, the magnetization is saturated and points OOP.[22] In such conditions, the orientation of the IP component is degenerate given that its magnitude is zero up to that small tilt. Therefore, it is plausible that a very small IP field is sufficient to break the degeneracy and reverse the magnetization. Nevertheless, we cannot rule out other coupling mechanisms between the IP magnetization component and the OOP, justifying further investigation.

Interestingly, an excursion at slightly lower fields (0.8≤$\mu_0 H_z^{exc}$≤2 T) only reverts the magnetization in part of the sample, as we can see for *N=1*,3, resulting in the formation of magnetic domains (**Fig. 3b,e**). **Figure 3e** presents a plot of the $B_z(x,y)$ image at $\mu_0 H_z$=0 for *N*=3, and reveals three magnetic domains in alternative orientations. Between the magnetic domains, we obtain domain walls with zero stray fields. The width of the domain walls is smaller than our spatial resolution. The magnetic signal measured for *N*=1 is a factor two larger than *N*=3. This difference is explained by the difference in tip-to-sample distance, which was 50 and 200 nm for *N*=1 and 3, respectively. The observed domain size varied from 100 nm to 2.5 μm for *N*=1 (see **Supplementary Figure 6**). All observed domains exhibit a magnetization parallel to the *b* axis. These domains are observed for odd layers down to the monolayer (**Fig. 3b**). In contrast, there are no domains in even layer samples, suggesting that the magnetic domains in these cases may not be stable or might be undetectable because of the AFM compensation between the layers and across different domains. For the same reason, we cannot determine whether the domain wall exists in the same location through all the layers in odd layer (*N*=3) samples or whether it is associated with the single uncompensated layer.

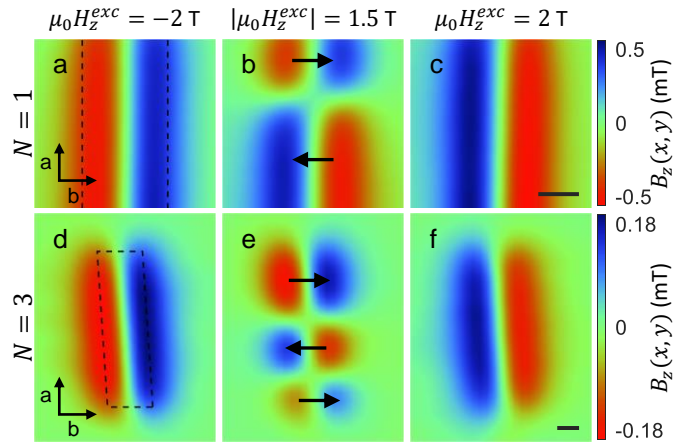

**Figure 3 Magnetic domains in CrSBr.** (**a-f**) SOT images of the out-of-plane component of the magnetic field $B_z(x,y)$ of magnetic domains in CrSBr with *N*=1 (**a-c**), and *N*=3 (**d-f**). The images are acquired at zero applied field after field excursion of $\mu_0 H_z^{exc}$=-2 T **a**, 1.5 T **b** 2 T **c**, -2 T **d**, -1.5 T **e**, and 2 T **f**. Black arrows indicate the direction of the magnetic moment of each domain parallel to the *b*-axis. The scale bar is 200 nm for all images. The red to blue color scale represents lower and higher magnetic fields, respectively, with a shared scale for each flake. The tip-to-sample distance was 50 nm (**a-c**) and 200 nm (**d-f**), which explain the difference in $B_z$ magnitude.

To address these questions, we have undertaken atomistic spin dynamic simulations (see Experimental Section for details) for a 100×100 nm$^2$ and $N$=2 system. As initial conditions, we build an atomically sharp domain wall at the center of the sample by setting the magnetization component for the top layer as follows: $M_a$=$M_c$=0, $M_b(a<L/2)$=1 and $M_b(a>L/2)$=-1, where $L$=100 nm is the sample dimension of the sample along the $a$-axis. The bottom layer is set with opposite magnetization to preserve the type-A AFM order ($M_a$= $M_c$=0, $M_b(a<L/2)$=-1 and $M_b(a>L/2)$=1). After letting the spins to thermally equilibrate for 10 million Monte-Carlo steps using the Metropolis algorithm framework,[33,34] we obtain the stable magnetization configuration resolved in space for each of the layers (**Fig. 4a-f**). We note the following observations. First, $M_c$ remains nearly zero for both layers and the spin rotates in the $ab$ plane as illustrated schematically in **Fig. 4j**. Second, we observe that the domain wall is of the order of 10 nm, which is smaller than our SQUID loop diameter, but consistent with the experimental data (**Fig. 3b,e**). Finally, we note that the interlayer AFM order is preserved, meaning that the sum of the magnetization for both layers ($\mu_{tot}$) is zero for each magnetization component. The vanishing $\mu_{tot}$ at the domain wall and its surroundings explains the absence of experimental observation of magnetic domain walls for samples with an even number of layers.

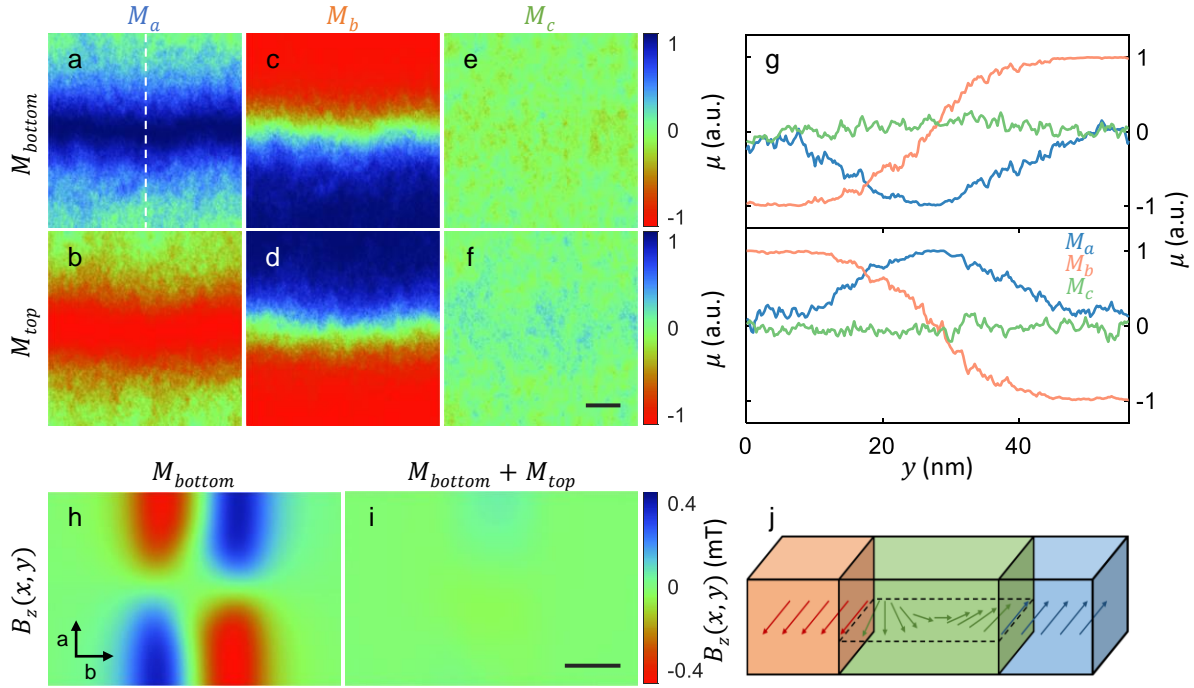

**Figure 4. Domain wall atomistic simulation for $N$=2 CrSBr.** (**a-f**) Spatial distribution of the magnetization component along $a$-axis **a,b**, $b$-axis **c,d**, and $c$-axis **e,f** calculated for the bottom **a,c,e** and top layer **b,d,f**. The simulations include dipolar interactions, Dzyaloshinskii-Moriya interaction, apart from exchange and anisotropic terms. They were conducted under zero applied magnetic field conditions and at a finite temperature of $T$ = 2 K. The scale bar is 10 nm. (**g**) Line cut from **a-f**, as indicated by the dashed line at **a**. (**h,i**) The OOP stray field $B_z(x,y)$ emanating from the domain wall by considering the magnetization distribution obtained for the bottom layer **h** and for the sum of the magnetization in the bottom and top layers **i**. The OOP stray field was calculated at a distance of 50 nm from the surface of the simulated sample. The scale bar is 100 nm. (**j**) A schematic illustration of the Néel domain wall found in CrSBr.

There are usually two types of domain walls in ferromagnets, namely Néel and Bloch, where the spins in the former remain in the plane (**Fig. 4j**) but rotate out of the plane in the latter. The atomistic simulations show that each layer contains a Néel-like domain wall. To verify that this is consistent with our SOT image, we calculate the OOP stray field $B_z(x,y)$ emanating from the domain wall by considering the magnetization distribution obtained for a monolayer (the bottom layer, **Fig. 4h**) and for a bilayer (both bottom and top layers, **Fig. 4i**). For the monolayer, we obtain the same characteristic vanishing $B_z(x,y)$ line at the domain wall that would not appear if we had OOP magnetic moment at a Bloch domain wall. For the bilayer, we obtain $B_z(x,y)$~0 as expected for $\mu_{tot}$=0. The compensation between the layers across the domain wall observed in the simulations explains why SOT did not detect them for even layers. The results for *N*=2 suggest that aligned domain walls is a stable configuration.

## Discussion

Our findings provide the first direct evidence of magnetic ordering and field-induced magnetic domains in CrSBr down to the 2D limit. Notably, it has previously only been possible to detect these magnetic properties indirectly, for example by transport or optical techniques.[7,17,21,22] Now, our experimental evidence, supported by atomistic simulations, demonstrate that the magnetic anisotropy and the magnetic moment density are independent of the sample thickness and are in good agreement with the bulk value. In general, the magnetic anisotropy is affected by the shape and magnetocrystalline anisotropy, where the former is determined by the aspect ratio of the sample and the resulting dipolar field, while the latter depends on the direction of the bounds participating in the magnetic exchange. Since shape anisotropy tends to orientate the magnetization along the longest axis of the sample, we would expect higher anisotropy in thinner films. In practice, DFT calculations predict a two-fold increase in the overall anisotropy for the monolayer compared to the bulk.[18,19] However, such simulations calculate the combination of the two factors that contribute to the anisotropy. In our case, the reduction of the magnetocrystalline anisotropy in the monolayer is balanced by the increase in the shape anisotropy, to keep the overall anisotropy nearly constant over all thicknesses. Finally, the anisotropy can explain the Néel domain wall since a Bloch domain wall would imply spins pointing along the hard OOP *c*-axis.

In conclusion, we report direct measurement of the magnetic texture in CrSBr down to the monolayer by nano SQUID-on-tip microscopy. Our results indicate that the magnetic anisotropy is surprisingly stable down to the monolayer. This result is consistent with the observation of Néel-like domain walls for all odd layers and demonstrates the importance of confinement on the magnetic anisotropy. These results provide valuable insights into the fundamental physics of magnetic materials, with potential implications for spintronics and data storage technologies such as atomic scale domain wall racetrack memories.[12]

## Experimental Section

**Sample fabrication:**

Crystal Growth: High-quality crystals of CrSBr were grown by solid state techniques, as previously reported.[22] The crystal structure was verified by powder and single crystal X-ray diffraction as well as by selected area electron diffraction in atomically-thin layers together with the elemental composition by energy-dispersive X-ray spectroscopy (EDS).[22] CrSBr atomically thin layers were obtained by the mechanical exfoliation of the bulk counterpart and placed on 285 nm $SiO_2$/Si substrates. The different materials were transferred onto the desired substrates for the SOT measurements using a deterministic transfer method with polycarbonate, as reported by Wang et al.[31] The entire process was performed inside an argon glove box to avoid any possible degradation. Samples, *N*=2,4 were encapsulated with hBN from both sides and *N*=1,3,5 were encapsulated with hBN from the top. In addition, one monolayer that was not encapsulated in order to determine the air stability was examined.

**Scanning SQUID-On-Tip microscopy:**

The SOT was fabricated using self-aligned three-step thermal deposition of Pb at cryogenic temperatures, as described previously.[26] All measurements were performed at 4.2 K in low pressure He (1 to 10 mbar). All images were acquired with a constant distance between the tip and sample (50 - 300 nm). Notably, the magnetic signal measured under these conditions was much larger than any possible influence of the varying topography, where the parasitic signal was estimated to be < 0.01 mT.

**Sample characterization:**

The thickness of samples *N*=2,4,5 was measured by scanning transmission electron microscopy (STEM) (**Supplementary Figure 7**). High-resolution scanning electron microscope cross-section lamellas were prepared and imaged by Helios Nanolab 460F1 Lite focused ion beam (FIB) - Thermo Fisher Scientific. The site-specific thin lamella was extracted from the CrSBr patterns using FIB lift-out techniques.[32] STEM and EDS analyses were conducted using an Aberration Probe-Corrected S/TEM Themis Z G3 (Thermo Fisher Scientific) operated at 300 kV and equipped with a high-angle annular dark field detector (Fischione Instruments) and a Super-X EDS detection system (Thermo Fisher Scientific). The thickness of samples *N*=1,3 was measured by optical contrast, based on the previous calibration performed with atomic force microscopy, Raman spectroscopy, and magneto-transport measurements.[22]

**Atomistic spin dynamics methods:**

Atomistic spin dynamics simulations were performed based on the VAMPIRE software package,[33] which has proven efficiency when investigating the presence and characteristics of non-trivial magnetic solitons in 2D-based materials.[34–42] Such approach was crucial to provide a theoretical benchmark that allows to obtain physical insights about the behavior of the spin inhomogeneous transition in the non-experimentally accessible potential domain walls that can be stabilized in AFM CrSBr-based bilayer samples. To model this system, we have employed a configurational spin Hamiltonian, $\mathcal{H}$, was employed given by

$$\mathcal{H} = -\sum_{i<j} \mathcal{J}_{ij}(\mathbf{M}_i \cdot \mathbf{M}_j) + \sum_{i<j} \mathbf{D}_{ij} \cdot (\mathbf{M}_i \times \mathbf{M}_j) + \mathcal{H}_{ani} + \mathcal{H}_{dip} \qquad (1)$$

where the sum runs over all the Cr atoms with magnetization vectors $\mathbf{M}_i$ and $\mathbf{M}_j$ at the *i*- and *j-th* atomic sites, respectively. Among these non-on-site interactions, one can find the symmetric Heisenberg exchange (which corresponds to the first term on the right-hand side of Eq. (1)), which was mediated by the exchange tensor, $\mathcal{J}_{ij}$ between Cr-based atomic sites, the Dzyaloshinskii-Moriya (DM) antisymmetric exchange (represented by the second input), being $\mathbf{D}_{ij}$ the Dzyaloshinskii vector, and the dipole-dipole contribution encapsulated in $\mathcal{H}_{dip}$. In this regard, it is important to note that the value of the FM-mediated intralayer exchange term $\mathcal{J}_2$ had been taken from Ref. [42] as a reference to define the magnitude of the $\mathcal{J}_{ij}$ components in order to obtain satisfactory predictions of the critical temperature of CrSBr-based systems.[42] On the other hand, the relative values between exchange parameters had been taken from Ref. [19]. All this information is summarized in **Supplementary Table 1** in **Supplementary Information**. Due to the absence of inversion symmetry between interacting Cr-based atoms,[43] the DM contribution was included, second term in Eq. (1), with DM unit vectors parallel to the *a-th* (mediating $\mathcal{J}_3$) and *b-th* (mediating $\mathcal{J}_1$) axes, being given by $D_1$=0.07 and $D_3$=0.18 meV, respectively.[19] Moreover, the long-range dipole-dipole interaction, $\mathcal{H}_{dip}$, which can be expressed as

$$\mathcal{H}_{dip} = -\frac{\mu_0 \mu_s^2}{4\pi} \sum_{i \neq j} \frac{3(\mathbf{M}_i \cdot \hat{\mathbf{r}}_{ij})(\mathbf{M}_j \cdot \hat{\mathbf{r}}_{ij}) - (\mathbf{M}_i \cdot \mathbf{M}_j)}{|\mathbf{r}_{ij}|^3} \qquad (2)$$

had been included for completeness despite the fact that the authors focused on the investigation of a purely AFM sample, using the tensorial approach for it,[33,44,45] being $\hat{\mathbf{r}}_{ij}$ the unit vector position joining the *i-* and *j-th* magnetic atoms. It should be noted that, in the previous expression, in addition to the presence of the vacuum magnetic permeability, $\mu_0$, the atomic magnetic moment, $\mu_s$, was involved, to which the value $\mu_s$=2.88$\mu_B$ has been assigned in consonance with the bulk scenario,[19] being $\mu_B$ the Bohr magneton.

Additionally, it was possible to identify in the spin Hamiltonian, given by Eq. (1), the on-site anisotropy (third term of the right-hand side), $\mathcal{H}_{ani}$, which had two uniaxial single-ion-based easy-axis terms, being represented by

$$\mathcal{H}_{ani} = -K_a \sum_i (\mathbf{M}_i \cdot \hat{\mathbf{a}})^2 - K_b \sum_i (\mathbf{M}_i \cdot \hat{\mathbf{b}})^2 \qquad (3)$$

where the values of the anisotropy constants, $K_a$ and $K_b$, govern the intermediate *a-* and easy *b-th* axes of the system, being given by $K_a$=8.06 µeV and $K_b$=31.53 µeV.[42] It is important to note that the previously introduced single-ion anisotropies were not, theoretically, the only ones that should contribute to the overall magnetocrystalline anisotropy of the system. The larger spin-orbit coupling of the Br atoms compared to those of Cr points to the existence of in-basal plane-based exchange anisotropy terms at the previously defined spin Hamiltonian.[46,47] However, in the computational characterization of the CrSBr-based system, the authors chose not to include these two-ion anisotropy terms, since the single-ion ones, characterized by Eq. (3), were enough to unravel the main features observed experimentally. Apart from this, in the literature there were two opposing positions about the two-ion contributions, since, on the one hand, it had been reported that they share the same order of magnitude as the on-site ones, inducing that the *a-th* axis becomes the easiest one in the system,[42] while, on the other hand, it had been postulated that the exchange interactions were practically isotropic and do not contribute substantially to the magnetic anisotropy distribution of a bilayer system.[45] It should also be noted that the experimentally reported atomic lattice spacings given by $a = 3.50$ and $b = 4.76$ Å were employed, taking into account that the height of the primitive cell is $c = 7.96$ Å.[19,20]

**Statistical Analysis**

The dominant source of systematic uncertainty in the evaluation of the magnetic moment was related to the tip-to-sample distance, which was measured to a precision of ±30 nm by coupling the SOT to a tuning fork and retracting by a known safe amount (50 to 300 nm). The uncertainty of this value depended on the calibration of the piezo scanner ($\approx$10%). In addition, the distance at which the tuning fork detects the surface was not precisely known ($\approx$20 nm above the surface). To estimate how this uncertainty affects the total magnetic moment $\mu_{tot}$, the SOT images were fit at different heights and the differences were noted in the resultant magnetic moment. Typically, the resulting uncertainty was of the order of 10%. The uncertainty of the angle θ, which was estimated to ±4° by comparing the result of the fit for two images acquired in the same conditions. The magnetic anisotropy constant depended on the θ and $\mu_{tot}$. The uncertainty propagation on $K_e$ for these parameters was calculated and a typical uncertainty of 30% was obtained. The simulation data displayed in **Figure 2** includes only the data that fell within the standard deviation (SD) range of the fitting line, for clarity. On average, for the net magnetic moment density, the SD is 4.5 $\mu_B$ nm$^{-2}$, covering 72% of the data. Similarly, on average, the SD for the canting angle is 4°, and this SD range includes 76% of the data. However, it was noted that the fitting calculations were done for the entire dataset.

## Supporting Information

Additional experimental details and discussion such as magnetic profile of the SOT images for $N$=1,3 CrSBr layers, SOT images of field evolution for $N$=1 to 5 layers, details of the magnetic simulation used in this work, a direct comparison between an encapsulated monolayer and one exposed to air, SOT images of magnetic domains in the monolayer, STEM characterization of the number of layers, and details on the atomistic spin dynamics Monte Carlo simulation.

## Acknowledgements

This work was supported by the European Research Council (ERC) Foundation grant No. 802952 and the Israel Science Foundation (ISF) Grant No. 645/23. The international collaboration in this work was fostered by the EU-COST Action CA21144. C.B.-C., S.M.-V and E.C. acknowledge financial support from the European Union (ERC AdG Mol-2D 788222 and FET OPEN SINFONIA 964396), the Spanish MCIN (2D-HETEROS PID2020-117152RB-100, cofinanced by FEDER, the Excellence Unit "María de Maeztu" CEX2019-000919-M), the Generalitat Valenciana (PROMETEO Program, and the PO FEDER Program IDIFEDER/2018/061), as well as the Advanced Materials program (supported by MCIN, with funding from the European Union NextGenerationEU (PRTR-C17.I1) and Generalitat Valenciana). E.J.G.S. acknowledges computational resources through CIRRUS Tier-2 HPC Service (ec131 Cirrus Project) at EPCC (http://www.cirrus.ac.uk) funded by the University of Edinburgh and EPSRC (EP/P020267/1); ARCHER2 UK National Supercomputing Service via Project d429. E.J.G.S. acknowledges the EPSRC Open Fellowship (EP/T021578/1), and the Edinburgh-Rice Strategic Collaboration Awards for funding support. The authors would like to thank Amir Capua for fruitful discussions. The authors thank Atzmon Vakahi and Sergei Remennik for technical support. R.R.-E. would like to thank Sarah Jenkins for technical support on the simulations and helpful discussions.

## Author Contributions

Y.Z. and A.N. equally contributed to this work.
Y.A., E.C., A.N., S.M.-V., and Y.Z. conceived the experiment.
C.B.-C., and S.M.-V. synthesized the CrSBr crystals and fabricated the CrSBr devices.
Y.Z., A.N. and N.F. carried out the scanning SOT measurements.
Y.Z., and A.N. computed the micromagnetic simulation.
A.N., and Y.Z. characterized the CrSBr devices.
Y.Z. and A.N. analyzed the data.
Y.A. and A.N. constructed the scanning SOT microscope.
M.E.H. developed the SOT readout system.
R.R.-E., and E.J.G.S. provided theoretical and simulation support.
A.N., S.M.-V, Y.Z., E.C., and Y.A. wrote the paper with contributions from all authors.
Notes: The authors declare no competing financial interest.

# Supplementary Information

Magnetic imaging and domain nucleation in CrSBr down to the 2D limit.

*Yishay Zur[†,1,2], Avia Noah[†,*,1,2], Carla Boix-Constant[3], Samuel Mañas-Valero*[3], Nofar Fridman[1,2], Ricardo Rama-Eiroa[4,5], Martin E. Huber[6], Elton J. G. Santos*[5,7,4], Eugenio Coronado[3], and Yonathan Anahory*[1,2]*

[†] These authors contributed equally to this work.
[1] The Racah Institute of Physics, The Hebrew University, Jerusalem, 9190401, Israel
[2] Center for Nanoscience and Nanotechnology, Hebrew University of Jerusalem, Jerusalem, 91904, Israel
[3] Instituto de Ciencia Molecular (ICMol), Universitat de València, Catedrático José Beltrán 2, Paterna 46980, Spain
[4] Donostia International Physics Center (DIPC), 20018 Donostia-San Sebastián, Basque Country, Spain
[5] Institute for Condensed Matter Physics and Complex Systems, School of Physics and Astronomy, University of Edinburgh, Edinburgh, EH93FD, United Kingdom
[6] Departments of Physics and Electrical Engineering, University of Colorado Denver, Denver, CO 80217, USA
[7] Higgs Centre for Theoretical Physics, University of Edinburgh, Edinburgh EH93FD, United Kingdom

*Correspondences to: avia.noah@mail.huji.ac.il, Samuel.Manas@uv.es, esantos@ed.ac.uk, yonathan.anahory@mail.huji.ac.il

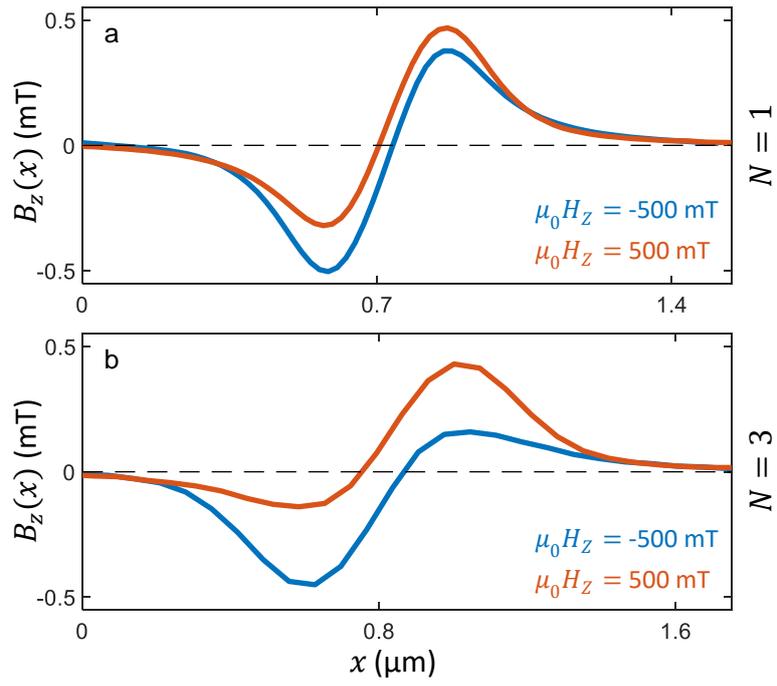

**Supplementary Figure 1. Magnetic profile of the images present in Figure 2 (a, b).** Magnetic profile of the images presented in Figure 2f,j for the monolayer **a**, and in Figure 2k,o for $N = 3$ layer **b**.

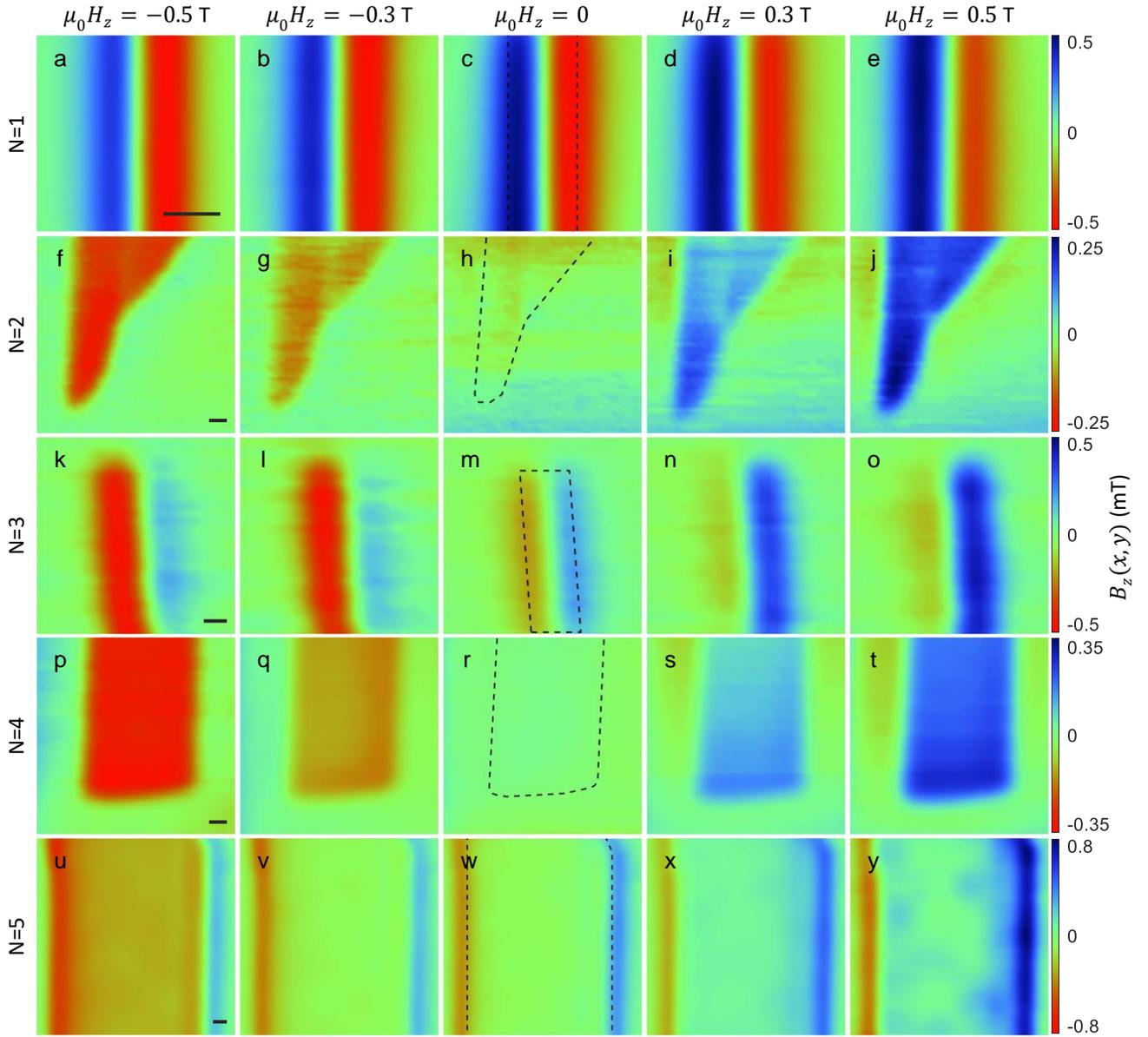

**Supplementary Figure 2. Field dependence.** (**a-y**) Sequence of magnetic SOT images of the OOP component of the local magnetic field $B_z(x,y)$ at $\mu_0 H_z = -0.5, -0.3, 0, 0.3, 0.5$ T after negative field excursion $|\mu_0 H_z^{exc}| > 1.5$ T for $N = 1$ **a-e**, $N = 2$ **f-j**, $N = 3$ **k-o**, $N = 4$ **p-t** and $N = 5$ **u-y**. The dash lines in **c, h, m, r, w** delineate the boundaries of the flake depicted in the corresponding row, while the scale bar in the left image of each row represents a size of 300 nm. The color scale is unique within each row.

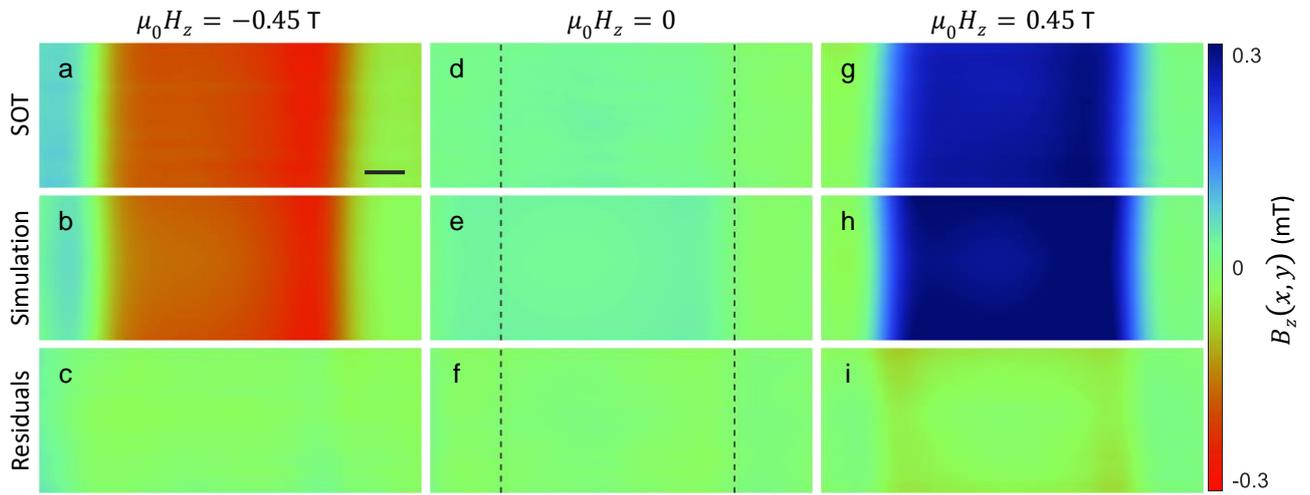

**Supplementary Figure 3. Simulation for even numbered of layers.** This figure presents a simulation of the magnetic SOT images of the OOP component of the local magnetic field $B_z(x,y)$ for a $N=4$ layers flake. The images were taken at $\mu_0 H_z = -0.45, 0, 0.45$ T (**a-c, d-f, g-i,** respectively) after a negative field excursion $\mu_0 H_z^{exc} = -2$ T. (**a, d, g**) SOT image. (**b, e, h**) Simulated image. (**c, f, i**) Residuals between the original SOT image and the simulated image. The dash lines in **d, e, f** indicate the flake edges. The scale bar in image **a** is 300 nm. The color bar is common to all images.

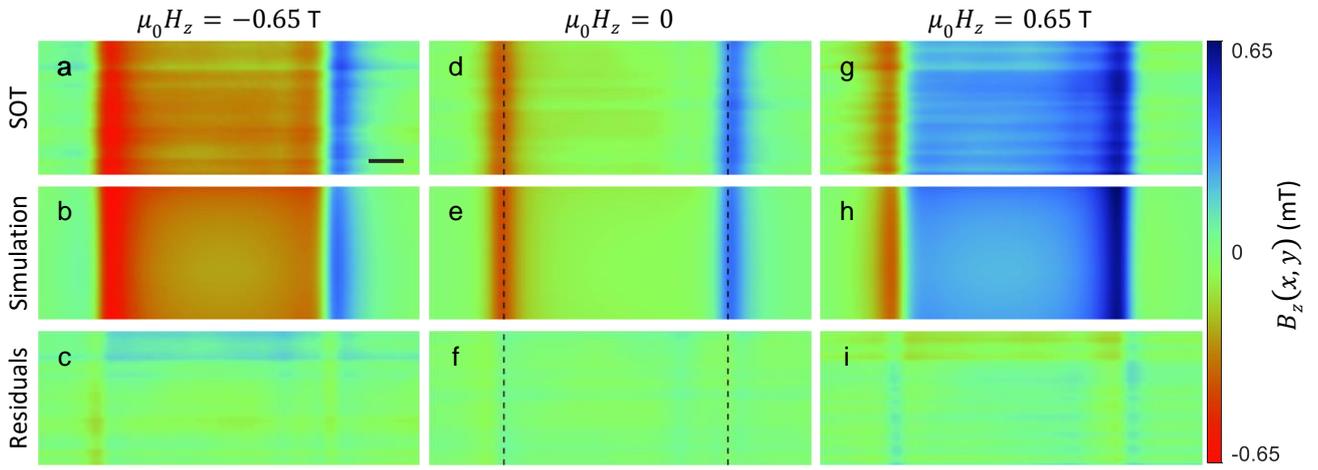

**Supplementary Figure 4. Simulation for odd numbered of layers.** Simulation of the magnetic SOT images of the OOP component of the local magnetic field $B_z(x,y)$ for $N = 5$ layers. The images were taken at $\mu_0 H_z = -0.65, 0, 0.65$ T (**a-c, d-f, g-i,** respectively) after a negative field excursion $\mu_0 H_z^{exc} = -1.5\ T$. (**a, d, g**) SOT image. (**b, e, h**) Simulated image. (**c, f, i**) Residuals between the original SOT image and the simulated image. The dash lines in **d, e, f** indicate the flake edges. The scale bar in image **a** is 500 nm. The color bar is common to all images.

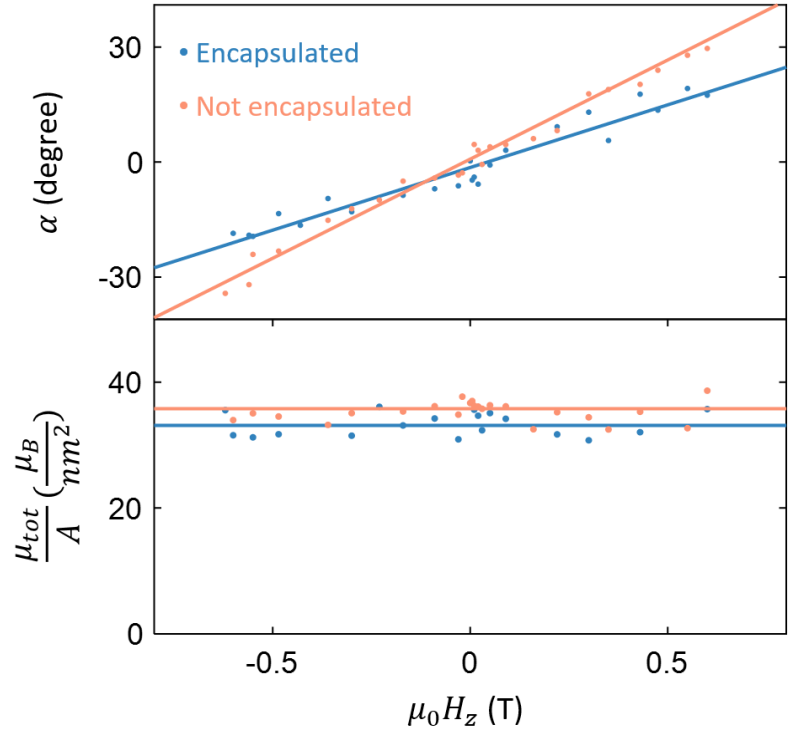

**Supplementary Figure 5. Anisotropy compared between encapsulated and a not encapsulated CrSBr monolayer.** Magnetostatic simulation result of the net magnetic moment density $\mu_{tot}$ and the polar angle $\alpha$ between the b and c axis. The anisotropy constant $K_e$ is similar for both cases.

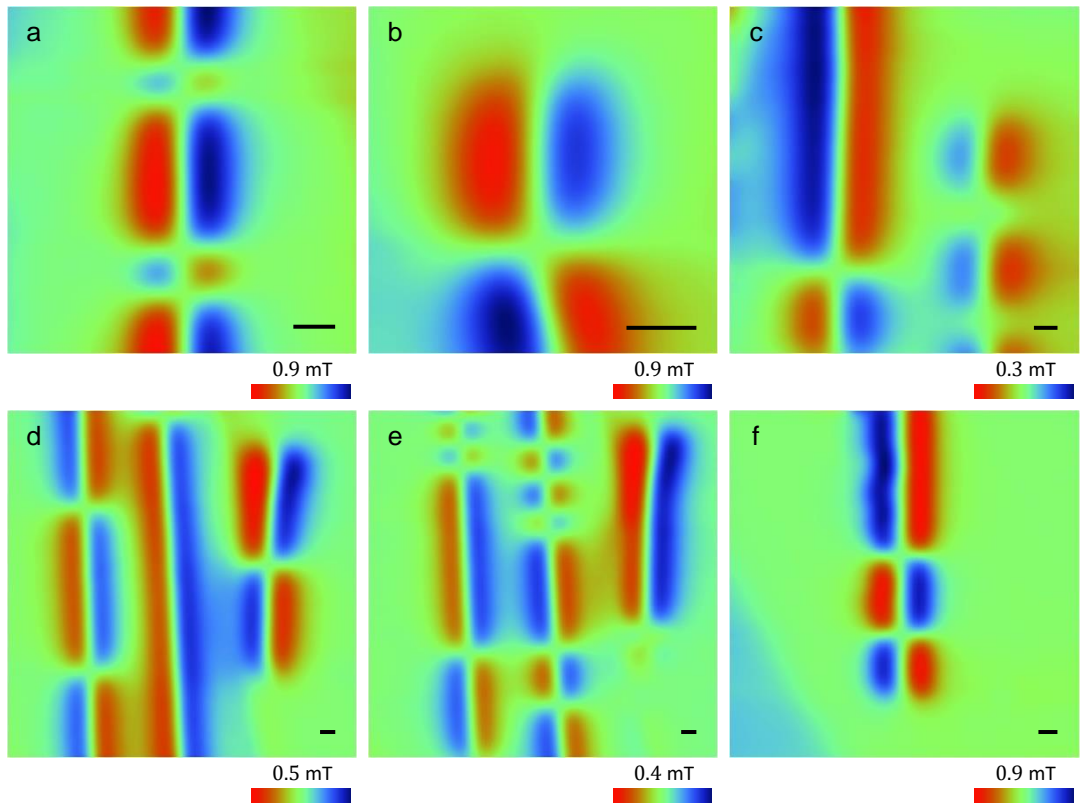

**Supplementary Figure 6. Magnetic Domains in monolayer.** Sequence of SOT images of magnetic domains of encapsulated CrSBr monolayer (**a-e**) and not encapsulated CrSBr monolayer (**f**). The tip-to-sample distance was 50 **a**, 50 **b**, 200 **c**, 200 **d**, 200 **e**, and 50 **f** nm. The scale bar in each image is 200 nm. The observed domain size varied from 100 nm (top domain at **a**) to 2.5 µm (left domain at **e**).

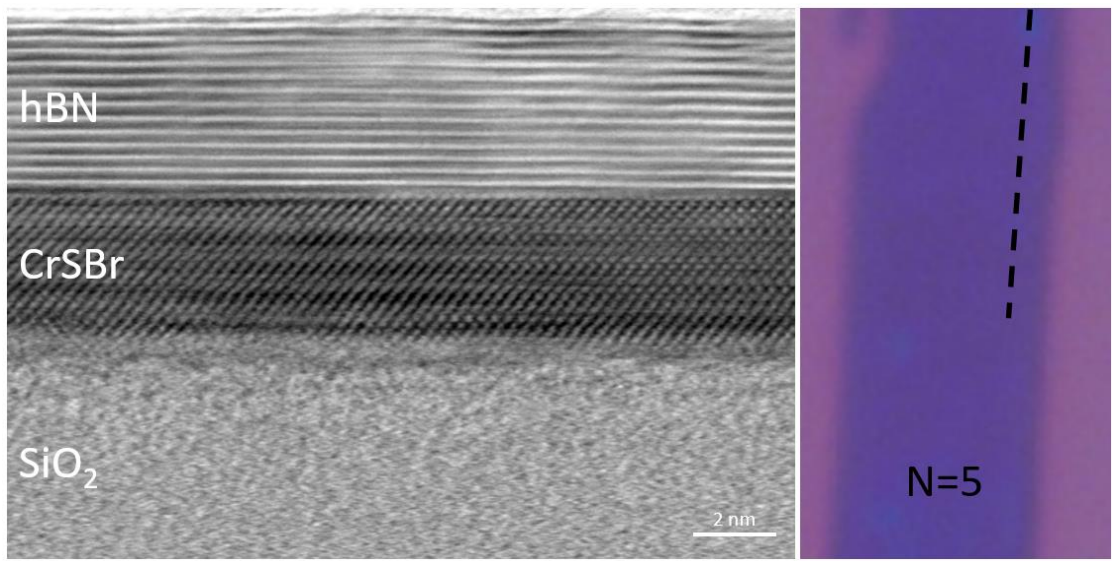

**Supplementary Figure 7. STEM Bright field image of the 5 layers flake.** (**a**) Cross-section of the CrSBr $N = 5$ layers flake on top of SiO$_2$ and encapsulated with hBN. (**b**) Optical image of the monolayer flake with a dashed line indicating where the cross-section was taken.

| Intramonolayer exchange interactions (meV) | |
|---|---|
| $\mathcal{J}_1$ | 7.2920 |
| $\mathcal{J}_2$ | 11.0800 |
| $\mathcal{J}_3$ | 4.4194 |
| $\mathcal{J}_4$ | -0.0032 |
| $\mathcal{J}_5$ | -0.0537 |
| $\mathcal{J}_6$ | -1.1995 |
| $\mathcal{J}_7$ | 0.4293 |
| **Intermonolayer exchange interactions (meV)** | |
| $\mathcal{J}_{z1}$ | -0.0025 |
| $\mathcal{J}_{z2}$ | 0.0025 |

**Supplementary Table 1.** Compendium of the intra and intermonolayer symmetric exchange contributions, $\mathcal{J}_i$, used in the atomistic spin dynamics simulations based on Eq. (1) of the main text.

Supplementary Note 1: Simulation details

To conduct our numerical calculations, we have studied the case of a CrSBr-based bilayer sample with in-basal-plane dimensions of 100 nm along the $a$- and $b$-th axes, and a thickness along the *c-th* spatial direction of 1.8 nm. In this regard, we have employed the experimentally reported atomic lattice spacings given by $a = 3.50$ and $b = 4.76$ Å, taking into account that the height of the primitive cell is $c = 7.96$ Å.[18,19] To explore the spin-based arrangement between the two uniformly magnetized spatial regions in an antiparallel fashion along the easy *b-th* axis, we artificially build an atomically sharp domain wall at the center of the sample ($L = 50$ nm along the *a-th* spatial direction) in each of the monolayers that make up the considered CrSBr-based bilayer sample. Initially, we set the magnetization components on the top monolayer such that $M_a, M_c(a < L/2 \ \& \ a > L/2) = 0$, $M_b(a < L/2) = 1$, and $M_b(a > L/2) = -1$, while in the bottom one we take into account the natural AFM behavior of a bilayer along the thickness, giving rise to $M_a, M_c(a < L/2 \ \& \ a > L/2) = 0$, $M_b(a < L/2) = -1$, and $M_b(a > L/2) = 1$. For a constant temperature of $T = 2$ K, in the absence of external stimuli, we have computed the evolution of the system for 10 million steps, using the Monte Carlo Metropolis algorithm,[31,45] to find the equilibrium configuration of the domain walls in each of the magnetic sublattices of the simulated structure. Additional simulations using Landau-Lifshitz-Gilbert (LLG) equation were used to describe the dynamics at different times and temperatures.